# Expanding IceCube GPU computing into the Clouds

Igor Sfiligoi
Univeristy of California San Diego
La Jolla, CA, USA
isfiligoi@sdsc.edu

Shava Smallen
Univeristy of California San Diego
La Jolla, CA, USA
ssmallen@sdsc.edu

Frank Würthwein
Univeristy of California San Diego
La Jolla, CA, USA
fkw@ucsd.edu

Nicole Wolter
Univeristy of California San Diego
La Jolla, CA, USA
nickel@sdsc.edu

David Schultz
University of Wisconsin - Madison
Madison, WI, USA
dschultz@icecube.wisc.edu

Benedikt Riedel
University of Wisconsin - Madison
Madison, WI, USA
briedel@icecube.wisc.edu

*Abstract*—The IceCube collaboration relies on GPU compute for many of its needs, including ray tracing simulation and machine learning activities. GPUs are however still a relatively scarce commodity in the scientific resource provider community, so we expanded the available resource pool with GPUs provisioned from the commercial Cloud providers. The provisioned resources were fully integrated into the normal IceCube workload management system through the Open Science Grid (OSG) infrastructure and used CloudBank for budget management. The result was an approximate doubling of GPU wall hours used by IceCube over a period of 2 weeks, adding over 3.1 fp32 EFLOP hours for a price tag of about $58k. This paper describes the setup used and the operational experience.

## I. ICECUBE COMPUTING

The IceCube Neutrino Observatory is the world's premier facility to detect high energy neutrinos and an essential part of multi-messenger astrophysics. Apart from the core data acquisition system, which is located at the South Pole, most of IceCube's compute needs are served by compute contributions from various research institutions, a large part of which uses the Open Science Grid (OSG) [1] infrastructure as the unifying glue.

IceCube has long been the leading user of GPU compute on the OSG, consuming over 80% of GPU hours available in both 2019 and 2020. Several IceCube codes run significantly faster on GPUs, including ray-tracing based detector simulation and machine learning codes. Unfortunately, GPU resources are still a relatively rare commodity in the scientific resource provider community, with only 8M GPU hours vs 550M CPU core hours available on OSG in 2020. This paper describes a two-week exercise aimed to validate the feasibility of expanding that resource pool with resources provisioned from commercial Cloud providers.

## II. INTEGRATING CLOUD RESOURCES INTO OSG

The OSG infrastructure is based on a federation principle, with each resource provider exposing a portal interface, also known as a Compute Element (CE) [2], and each user community then building an overlay workload management across them, typically using glideinWMS.

We thus instantiated a dedicated HTCondor-based CE, provisioning a dedicated Virtual Machine (VM), and registered it in OSG with the stated policy of only accepting IceCube jobs. On the backend, we created VM images containing the standard OSG client software [6] and then used the Cloud-native group provisioning mechanisms to start the worker VMs, also known as cloud *instances*.

We provisioned resources from all three main commercial Cloud providers, namely Microsoft Azure, Google Cloud Platform (GCP) and Amazon Web Services (AWS). In order to maximize the return on investment, we used only the smallest instances providing NVIDIA T4 GPUs, which we previously measured to deliver the best value for IceCube [3]. We also provisioned all VMs in the cheaper preemptible mode, also known as *spot instances*, since the OSG infrastructure can gracefully deal with preemption and past experiments showed that it is cost effective even at high scales.

All three Cloud providers offer group provisioning mechanisms with very similar semantics. We used Azure Virtual Machine Scale Sets (VMSS), GCP Instance Groups, and AWS Spot Fleets. All three allowed us to set the desired number of instances in a specific region, and they would provision as many as available at that point in time; no further operator intervention was needed. Since all Cloud providers have many independent regions, we would typically instantiate one group mechanism per region.

## III. ACCOUNT CREATION AND BUDGET MANAGEMENT

One major concern when executing multi-Cloud provisioning is budget management. While it is relatively easy to get spend data from the native Cloud interfaces, one would have to manually aggregate the information from the various providers and keep track of the remaining budget. We thus decided to partner with the CloudBank [4] project to have them manage our NSF-provided funds and track our spending rate.

CloudBank provides several budget reporting and management services, but for our purposes the two simplest ones provided all the needed functionality. The first one is a Web page providing a single window showing the total spending, both per provider and aggregate, the remaining budget and the fraction compared to the total budget. The other service is a

This work has been partially funded by the US National Science Foundation (NSF) Grants OAC-1941481, CNS-1925001, OAC-1841530, MPS-1148698, OAC-2030508 and OAC-1826967.

Pre-print version, July 2021.

periodic email, generated at periodic spending thresholds, e.g. less than 50% of the budget remaining, which provides both the remaining budget amount and fraction, and the spending rate over the past few days. Between the two, we had a good picture of how the budget situation was evolving and made provisioning corrections as appropriate.

Finally, we also used the CloudBank service for establishing a new account in one of the Cloud providers; we already had accounts in the other two and were able to simply link them with the CloudBank accounting system. While establishing a new Cloud account for personal use is trivial, it can be surprisingly complicated in an institutional environment, both from an accounting and legal standpoint. CloudBank is uniquely positioned in making this process very simple for any NSF-funded activity.

## IV. Operational Experience

Apart from planning, the whole exercise was executed in a period of just over two weeks. After deploying the CE and registering it with OSG, we initially provisioned a small number of VMs in each of the targeted Cloud regions to validate the setup. The initial validation did not expose any major issues in either AWS or GCP, but did bring to the surface the need for frequent TCP alive messages in Azure. The default Azure NAT setup has a 4-minute timeout on idle outgoing TCP connections, which HTCondor uses for job management purposes, and the default OSG setup was set to 5 minutes, resulting in constant preemption of the user jobs. Once that parameter was adjusted, all regions in all three Cloud providers were successfully executing user jobs to completion.

We spent the next few days slowly raising the number of instances in each of the targeted Cloud regions and monitoring the preemption rate. We were pleasantly surprised to find Azure, who has the lowest prices for spot T4 GPUs at $2.9/T4 day, to have plenty of spare capacity with very low preemption rates. We thus heavily favored Azure during most of the exercise.

After we were confident that the setup was sound, we ramped up in steps to 400, 900, 1.2k, 1.6k and finally to 2k GPUs, sustaining at each step for extended periods of time to validate the stability of the system before moving higher. We did not experience any intrinsic system problems during the whole period, although at one point one of the IceCube users filled up his disk area, resulting in a partial degradation of the system, from which it auto-recovered.

Moreover, just as we were running using 2k GPUs, the Cloud provider hosting the CE had a major network outage, resulting in the total collapse of the backend workload management system. We quickly de-provisioned all the worker instances, by instructing the various Cloud-native group mechanisms to keep zero active instances, so there was minimal financial loss involved. The situation was resolved after a couple of hours, and we resumed operation for a few more days at a lower instance count of 1k GPUs, since at that point in time we had only about 20% of the budget left. A snapshot of the IceCube monitoring is available in Fig. 1.

## V. Summary and Conclusion

The exercise of adding Cloud-provisioned GPU capacity to the IceCube environment was very straightforward. By abstracting the Cloud details behind an OSG CE and configuring the Cloud instances with standard OSG software, IceCube users' jobs executed exactly the same way as they would on any on-prem resource. Moreover, while we did limit the access to the IceCube community, the same exact setup could have been used to serve any other set of OSG communities, too.

No unusual problems were encountered during the two-week period, during which time we more than doubled the number of GPU hours that IceCube had access to, as shown in Fig. 2. The total cost of this exercise was approximately $58k, all included, which allowed us to deliver 16k GPU days or about 3.1 fp32 EFLOP hours of compute.

Using CloudBank for account creation and budget management further simplified the endeavor. The human cost of operation was also modest and mostly directed toward pushing the limits of the setup, due to the experimental nature of the Cloud run. In future work, we will engage with other scientific communities who are interested in leveraging our infrastructure.

## Acknowledgment

This work has been partially funded by the US National Science Foundation (NSF) Grants OAC-1941481, CNS-1925001, OAC-1841530, MPS-1148698, OAC-2030508 and OAC-1826967.

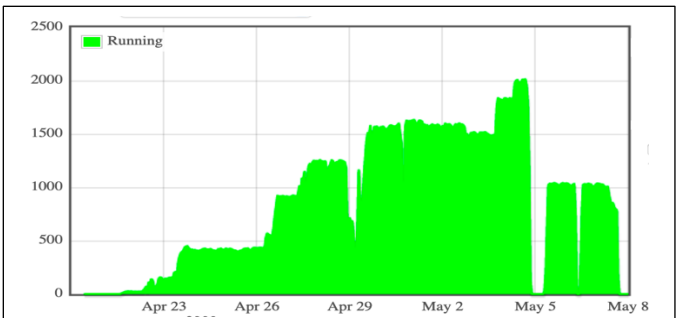

Fig. 1: Snapshot of the Cloud run as seen by IceCube users

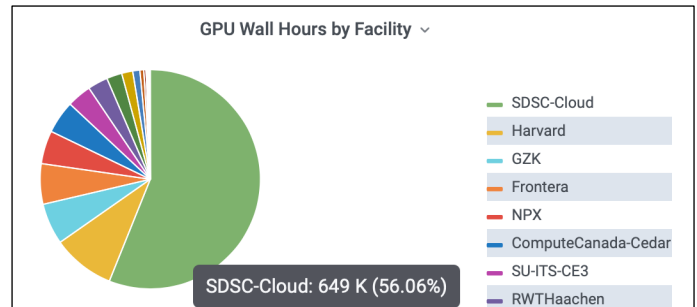

Fig. 2: Snapshot of OSG accounting for IceCube during the same period